\documentclass[12pt]{article}
\usepackage{epsf}
\usepackage{graphicx}
\usepackage{axodraw}
\usepackage{psfig,color,epsfig}
\usepackage{eepic}
\usepackage{latexsym}
\newcommand{\be}{\begin{equation}}
\newcommand{\ee}{\end{equation}}
\newcommand{\ba}{\begin{array}}
\newcommand{\ea}{\end{array}}
\newcommand{\bqa}{\begin{eqnarray}}
\newcommand{\eqa}{\end{eqnarray}}

\begin{document}

\begin{center}
{\Large\bf A unitarity analysis on the I=0 $d$ wave $\pi\pi$
scattering amplitude }

\vspace{10mm}
 {\it  {\sc  Jian-jun Wang,\footnote{Present address: Chinasoft Resource Corporation,
 No.4285 Zhichun Road, Haidian District, Beijing, 100080 P.R. China
} Z.~Y.~Zhou and H.~Q.~Zheng }
\\[5mm]
{Department of Physics, Peking University, Beijing 100871,
P.~R.~China }
\\[5mm] }

\today
\begin{abstract}
We study I=0 $d$ wave $\pi\pi$ scattering phase shift using a
proper unitarization approach. It is verified that the $f_2(1270)$
resonance corresponds to a twin pole structure: one on the second
sheet, another one on the   third sheet. Also we find fair
agreement on $f_2(1270)$ pole's mass and width between our results
and the PDG values. Besides, our analysis reveals the existence of
a virtual state in this channel. Its pole location is determined
by fitting experimental data and it is found to be in good
agreement with the prediction from chiral perturbation theory. Our
analysis demonstrates that partial wave amplitudes calculated in
chiral perturbation theory are reliable  in the small $|s|$
region, contrary to some unitarization models.
\end{abstract}
\end{center}

\vspace{1cm}

In a series of recent publications we have developed a new
dispersion approach incorporating unitarity and chiral symmetry,
to study scatterings between pseudo-Goldstone bosons. The approach
is found to be particularly useful in revealing low energy pole
structure of the scattering amplitudes. It is demonstrated that
the $\sigma$ meson is essential to adjust chiral perturbation
theory ($\chi$PT) to experimental data.~\cite{XZ00} Combining with
crossing symmetry, the mass and width of the $\sigma$ meson are
carefully determined~\cite{zhou05} and are found in nice agreement
with the Roy equation analysis.~\cite{CGL01} By analyzing the LASS
data~\cite{LASS} it is also demonstrated that the $\kappa$
resonance must exist in I=0,J=1/2 channel of $\pi K$ scatterings
if $\chi$PT prediction on the scattering length in the
corresponding channel is acceptable.\cite{piK} It is also found
that, interestingly, in I=2,J=0 channel of $\pi\pi$ scattering, a
virtual state pole is needed in order to fit the experimental data
and to correctly reproduce the chiral prediction on the scattering
length.~\cite{zhou05} The existence of such a virtual state pole
is actually a prediction of current algebra and
$\chi$PT~\cite{ang}, hence demonstrating that $\chi$PT
  provides a consistent description to the scattering $T$
matrix in the small $|s|$ region. In this note we will further
investigate low energy I=0 $d$--wave $\pi\pi$ scatterings, using
the phase shift data from Ref.~\cite{hyams}. The results provide
further evidences to support $\chi$PT predictions in the small
$|s|$ region.

We start from a proper parametrization form,  for the partial wave
$S$ matrix~\cite{zhou05,Ztalk03}:
 \be\label{param}
S^{phy}=\prod_iS^{p_i}\cdot S^{cut}\ ,
 \ee
 where $S^{p_i}$ are
the simplest $S$ matrices characterizing the isolated
singularities of $S^{phy}$ on the second sheet in the absence of
bound states, which are:
\begin{enumerate}
\item For a virtual state pole  located  at $s=s_0$
($0<s_0<4m_\pi^2$),
  \be \label{vsp} S^{}(s)=\frac{1+i\rho(s)a }{1-i\rho(s)a}\
, \ee
 and $a$ is the scattering length:
 \bqa\label{v_s_l}
  a^{ }&=&\sqrt{s_0\over 4m_\pi^2-s_0}\ \ .
    \eqa
  The kinematic factor
$\rho(s)={{2k(s)}\over {\sqrt s}}=\sqrt{s-4 m^2_\pi\over s}\ .%
$
 \item For a  resonance located at $z_0$ (and
$z_0^*$) on the second sheet, the $S$ matrix can be written as,
  {\be\label{resp}
S^R(s)=\frac{M^2[z_0]-s+i\rho(s)s G[z_0]}{M^2[z_0]-s-i\rho(s)s
G[z_0]}\ , \ee } where { \bqa M^2[z_0]&=&{\rm Re}[z_0] + {\rm
Im}[z_0]\frac{{\rm Im}[\sqrt{z_0(z_0-4m_\pi^2)}]}{{\rm
Re}[\sqrt{z_0(z_0-4m_\pi^2)}]}\ \ , \nonumber \\
G[z_0]&=&\frac{{\rm Im}[z_0]}{{\rm Re}[\sqrt{z_0(z_0-4m_\pi^2)}]}\
. \eqa}
\end{enumerate}
According to Eq.~(\ref{param}), each pole's contribution to the
scattering phase shift, denoted as $\delta_{p_i}(s)$, is additive
and easily calculable using Eqs.~(\ref{vsp}) and (\ref{resp}). In
Eq.~(\ref{param}) $S^{cut}$ can be parameterized in the following
simple form,
 \bqa\label{fS'}\
 S^{cut}&=&e^{2i\rho f(s)}\ ,\\
  f(s) &\equiv&f_L(s)+f_R(s)= {s\over \pi }
{\int_L}{{\mathrm{Im_L}f(s')\over{(s'-s)s'}}ds'} + {s\over \pi}
{\int_R}{{\mathrm{Im_R}f(s')\over{(s'-s)s'}}ds'}\ ,\label{fS}
 \eqa
where $L=(-\infty,0]$ denotes the left hand cut on the negative
real axis,   $R$ denotes cuts from the first inelastic threshold
to $\infty$, and the dispersion integral is free from subtraction
constants. The discontinuity of $f(s)$ on left or right hand cut
is expressed as
  \be\label{discf}
 \mathrm{Im}_{L,R}f(s)=-{1\over
{2\rho(s)}}\log|S^{phy}(s)|\ .
  \ee
The above parametrization is, however, not directly applicable to
the study of $d$--wave scatterings since the following threshold
constraint has not yet been taken into account:
 \be\label{ab}
\mathrm{Re}T_{IJ}(s)=q^{2J}[a^I_J+b^I_Jq^2+O(q^4)]\ ,\,\,\,\,
(q=\frac{1}{2}\sqrt{s-4m_\pi^2}\,)\ , \ee
 To remedy this we first recast Eq.~(\ref{fS}) into a
thrice subtracted form with a once-subtraction at  $s=0$, and a
twice-subtraction at physical threshold $4 m^2_\pi$:
 \bqa f(s) &=& s\,\left( -f'(4m^2_\pi) +
     \frac{f(4m^2_\pi)}{2\,{{m_{\pi }}}^2} \right)
   + s^2\,\left( \frac{-f(4m^2_\pi)}
      {16\,{{m_{\pi }}}^4} +
     \frac{f'(4m^2_\pi)}{4\,{{m_{\pi }}}^2} \right)\nonumber \\%
 &+&{s(s-4m^2_\pi)^2\over \pi }
{\int_{L+R}}\,\,{{\mathrm{Im}f(s')\over{(s'-s)s'(s'-4m^2_\pi)^2}}ds'}
\ ,\label{fS0}\eqa
 where the two subtraction constants, $f(4m^2_\pi)$
and $f'(4m^2_\pi)$ are not free, they are correlated to pole
parameters as dictated by Eq.~(\ref{ab}):
\bqa\label{ab'} f(4m^2_\pi)&=& \sum_i f_{p_i}\ , \nonumber \\%
f'(4m^2_\pi)&=& \sum_i f'_{p_i}\ ,   \eqa
 where
\bqa f_r&=&\frac{4\,{G}\,{{m_{\pi }}}^2}{4\,{{m_{\pi }}}^2-{{M}}^2}\ , \nonumber \\%
f'_r&=&-\frac{ {G}\,\left( 16\,{{G}}^2\,{{m_{\pi }}}^4 +
12\,{{m_{\pi }}}^2\,{{M}}^2 - 3\,{{M}}^4 \right)  }
   {3\,{\left( 4\,{{m_{\pi }}}^2 - {{M}}^2 \right) }^3}\eqa for a
   resonance, and
\be f_{v}=-a \ ,\,\,\, f'_{v}=\frac{{{a}}^3}{12\,{{m_{\pi
}}}^2}\ee for a virtual state. Since the threshold constraint must
be exactly obeyed, we in the following will make use of
Eq.~(\ref{ab'}) to replace $f(4m_\pi^2)$ and $f'(4m_\pi^2)$ in
Eq.~(\ref{fS0}) by corresponding pole parameters.

 According to Eqs.~(\ref{param}), (\ref{fS'}) and
(\ref{fS}), the phase shift from partial wave $S$ matrices for
elastic $\pi\pi$ scatterings consist of contributions from
second-sheet poles, left--hand cut integral and right--hand cut
integral: \be\label{delta}
\delta(s)=\sum_i\delta_{p_i}^{\mathrm{II}}(s)+\rho(s)f(s) \ee The
right--hand cut represents effects from inelastic thresholds and
poles from higher sheets.  We have:~\cite{XZ05} \bqa\label{IMR}
&&\mathrm{Im}_{R}f(s)=-{1\over
{2\rho(s)}}\log|S^{phy}(s)|\nonumber\\
&&=
-\frac{1}{4\rho}\log\left[1-4\rho\mathrm{Im}_{R}T+4\rho^2|T(s)|^2\right]\,\nonumber\\
&&= -\frac{1}{4\rho}\log\left[1-4\rho(\sum_{n\neq
1}\rho_n|T_{1n}(s)|^2+\cdots )\right]\
\eqa%
As an approximation, when only poles are considered, we have the
following parametrization for inelastic amplitudes:\cite{collins}
  \bqa\label{inelasticT}T_{1n}=\frac1{\sqrt{\rho_1(s)\rho_n(s)}}
  \sum_{r}{ {\cal M}_r(\Gamma_{r1}\Gamma_{rn})^
  {\frac{1}{2}}\over {\cal M}_r^2-s-i{\cal M}_r \Gamma_r}\, ,
  \eqa
with $\Gamma_{rn}$ the partial width into channel $n$,
$\Gamma_{r}$ the total width of resonance $r$, and $\rho_n(s)$
represents the kinematic factor of $n$-th channel. Especially in
above (and hereafter) subscript 1 denotes the $2\pi$ channel, and
here $R$ starts from 4$\pi$ threshold. It should be stressed that
the Eq.~(\ref{inelasticT}) only works for narrow resonances which
is only of limited usage, and furthermore, the narrow width
approximation only works when the energy region is far from any
physical threshold.

\begin{figure}%
\begin{center}%
\vspace{2cm}
\mbox{\epsfxsize=70mm\epsffile{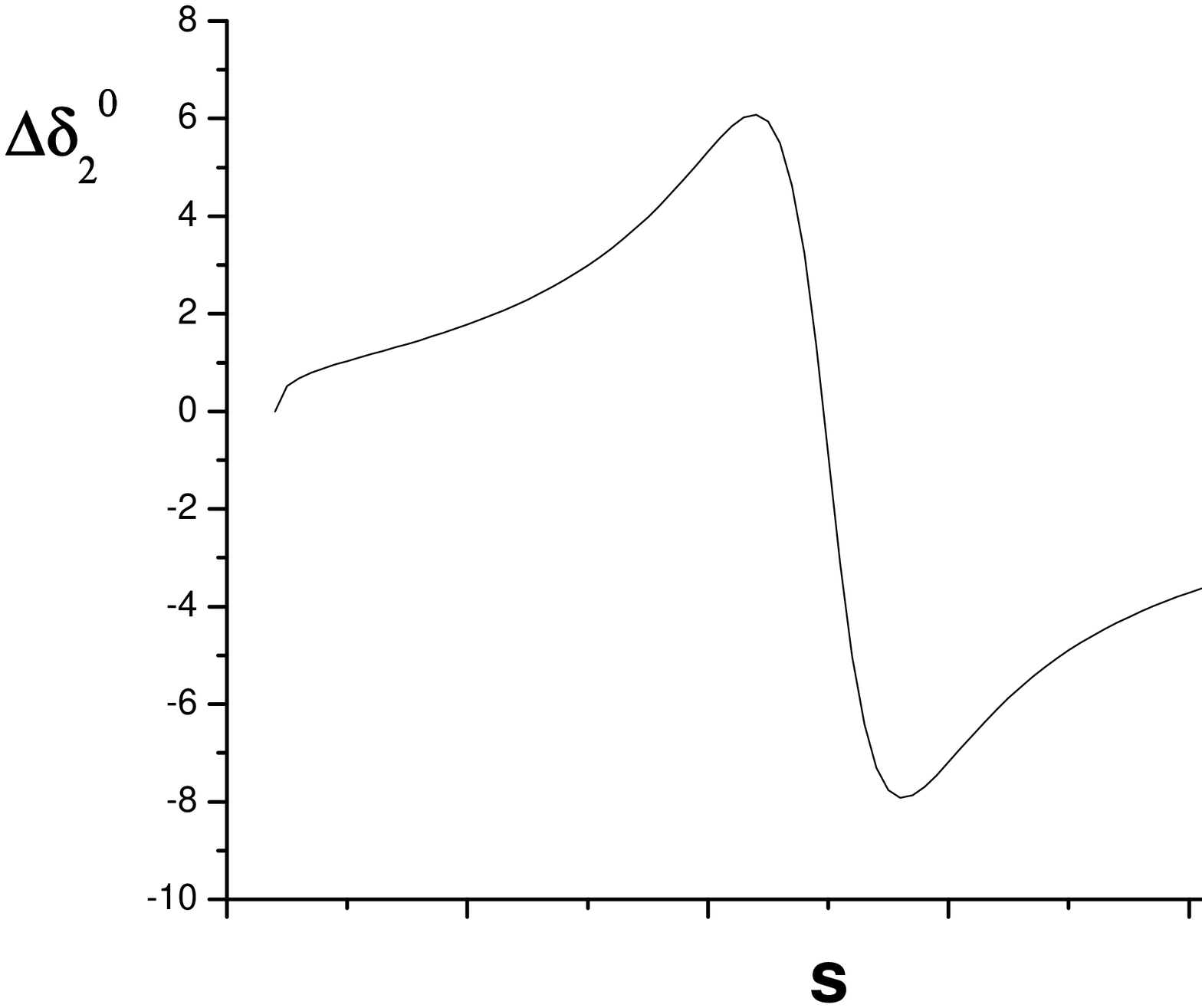}}%
\mbox{\epsfxsize=70mm\epsffile{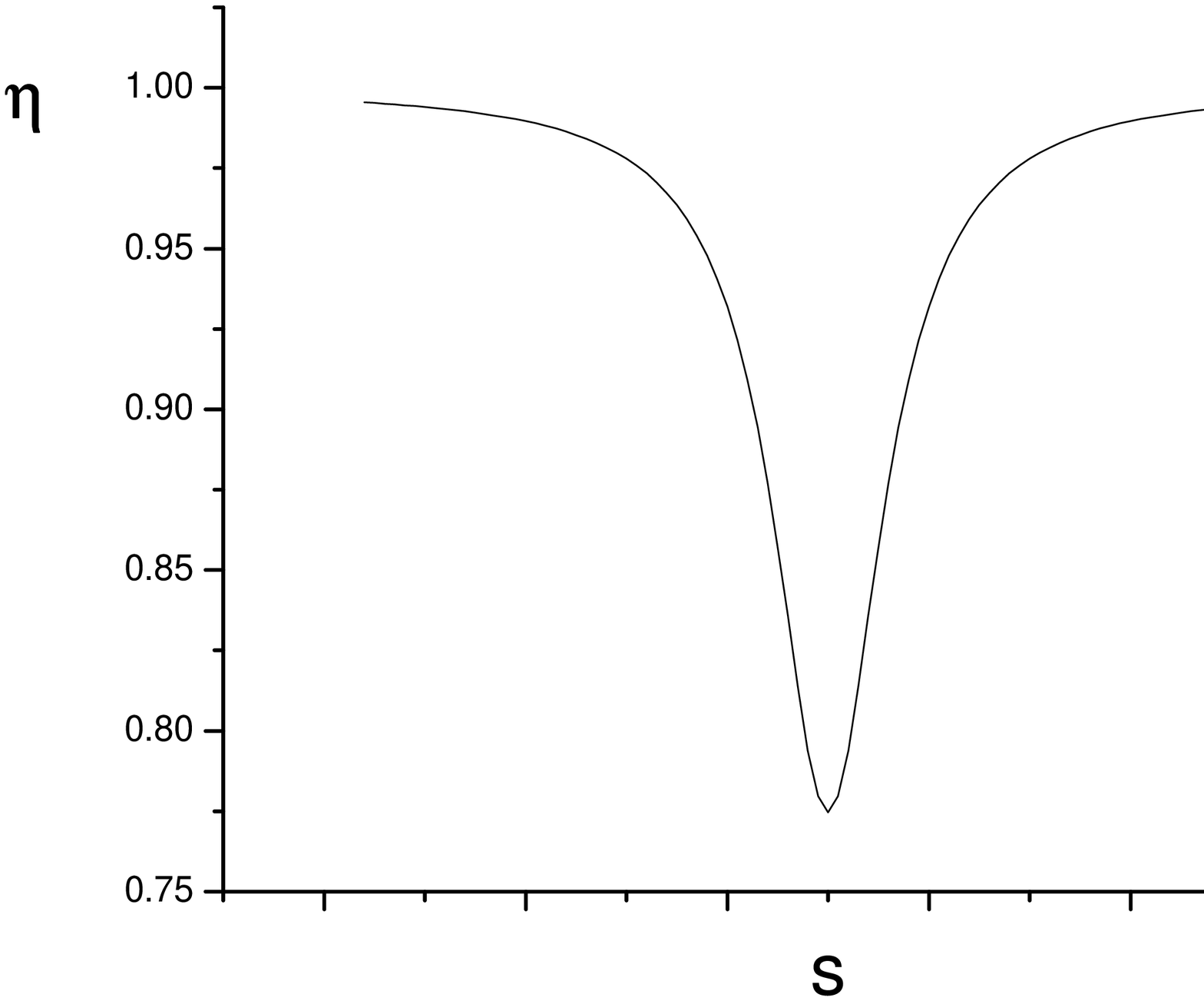}}%
\vspace{-2cm}
 \caption{\label{kinkanddip}  An illustrative example
for a single resonance's contribution (hence without interference
between different resonances) to the phase shift and the
inelasticity parameter above the in-elasticity threshold, drawn
using Eqs.~(\ref{delta1'}) and (\ref{eta1'}).}
\end{center}%
\end{figure}

\begin{figure}%
\begin{center}%
\includegraphics[width=0.9\textwidth]{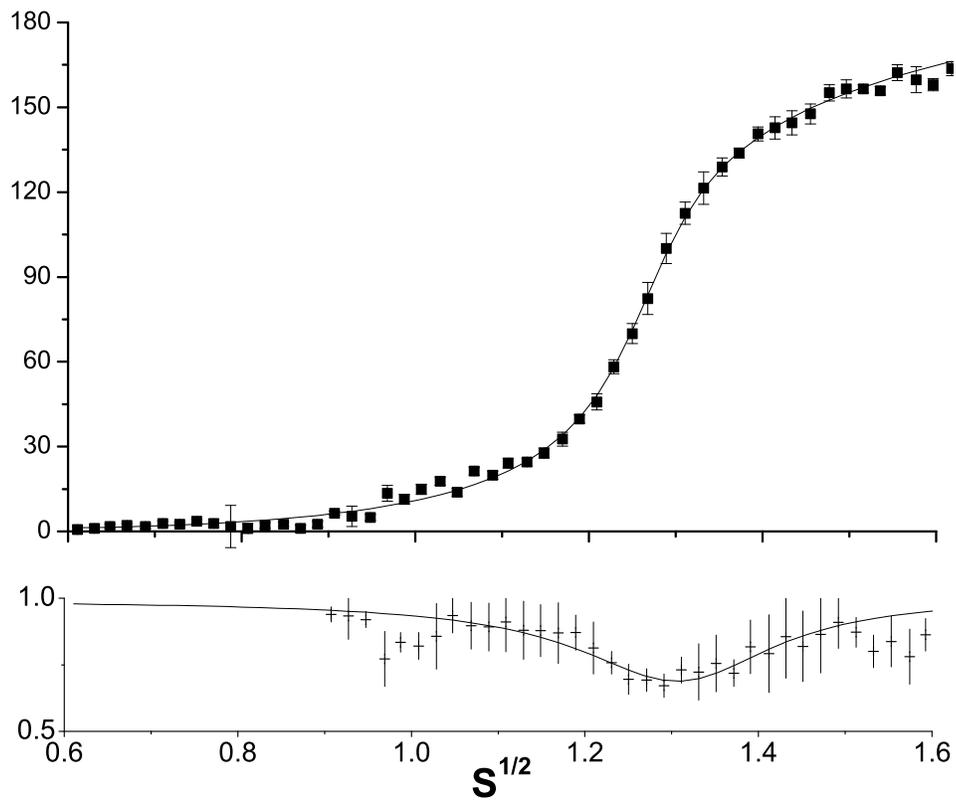}
\caption{\label{fitrhc} The fit on I=0 $d$-wave $\pi\pi$
scattering phase shift. Data are taken from
Ref.~\cite{hyams}. The higher resonances' effects are not included.}%
\end{center}%
\end{figure}

For the parametrization scheme described above, it should be
realized that isolated singularities in a couple channel $S$
matrix can play rather different roles depending on which sheet
they locate. It is therefore necessary to elaborate more about the
relation between a physical resonance and its pole structure in
the couple channel situation. Taking the well established
$f_2(1270)$ I=0 $2^{++}$ resonance~\cite{PDG04} for example, the
PDG value of its mass and width
 are $M=1275.4\pm 1.1$MeV and $\Gamma=185.2^{+3.1}_{-2.5}$MeV. The
 particle decays predominantly into $\pi\pi$ with a branching
 ratio $\mathrm{Br}({\pi\pi})=84.8\%$. The second largest branching
 ratio is the $4\pi$ mode, $\mathrm{Br}({4\pi })=10.2\%$ with
 sizable error bars. The $\bar KK$ mode is less important, with $\mathrm{Br}({\bar
 KK})=4.6\%$. Most of the experimental results listed in
 Ref.~\cite{PDG04} are from production experiments using a
 standard Breit--Wigner parametrization form for the $d$-wave,
 which means the detected $f_2(1270)$ resonance is on the sheet
 smoothly connected to the upper edge of the unitarity cut (i.e., the place where
 experimentalists perform experiments). In a more formal language, under for example
 the two channel
 approximation, the experimentally established resonance is a 3rd sheet pole. However, in the
 present situation the picture of a couple channel resonance is more
 complicated.
Indeed, phase shift data in the I,J=0,2 channel exhibit a typical
resonance structure with a mass around 1270MeV, as shown in
fig.~\ref{fitrhc}. Nevertheless it is easy to understand from
Eq.~(\ref{param}) that the jump of the scattering phase shift at
around 1270MeV is provided by a second sheet pole. In the present
scheme, the expression for the  phase shift of the elastic
scattering amplitude measured by experiments in the in-elastic
region, according to Eq.~(\ref{delta}), reads as
 \be\label{delta1'}
  \delta(s)=\sum_i\delta_{p_i}^{\mathrm{II}}+\rho_1 f_L(s)+{s\rho_1(s)\over \pi}\,{\cal P}
{\int_R}\,{\mathrm{Im_R}f(s')\over{(s'-s)s'}}ds'\ , \,\,\,
(s>16m_\pi^2)
   \ee
where ${\cal P}$ stands for principal value. The inelasticity
parameter obeys the following expression,
 \be\label{eta1'}
  \eta(s)=\exp[{-2\rho_1(s)\mathrm{Im}_Rf(s)}]\ .
   \ee
From Eqs.~(\ref{IMR}) and (\ref{inelasticT}) we know that
physically observed narrow states in production experiments
contribute
 to the inelasticity parameter.  Meanwhile it can be verified that
 their contributions to
 the phase shift of the elastic amplitude are small kinks, rather
 than  rapid jumps of approximately 180$^\circ$ (see
 fig.~\ref{kinkanddip} for illustration). In other words, if we could have accurate data
 on phase shift and especially inelasticity, we would be able to
 get more information on poles on different sheets.\footnote{More precisely speaking, under present scheme
 we can at best determine
 the partial decay width $\Gamma_1$ and the total decay width $\Gamma_{tot}$ but no more.} Nevertheless
 in practice the data are usually not good enough to achieve such a goal.

The left hand integral appeared in Eq.~(\ref{fS0}) converges very
rapidly if we approximate $S^{phy}$ in the integrand by $S^{\chi
PT}$. It has been shown that such an approximation provides a
satisfactory description to the data at qualitative level in
exotic channels, and also provides at least a self-consistent
description in non-exotic channels.~\cite{zhou05,piK,XZ00}
Therefore we in this paper will also use $\chi$PT result to
evaluate the left hand integral. Owing to the thrice-subtraction,
the cutoff parameter used when evaluating the left hand integral
has only a tiny influence, and increasing the cutoff parameter
will slightly improve the fit quality. Thus the cutoff parameter
is  practically fixed at $\infty$ and and is not treated as a free
parameter.

The integral on the $r.h.s.$ of Eq.~(\ref{fS0}) behaves as
$O(s^2)$ when $s$ is large. The coefficient of the $O(s^2)$ term
has a negative sign and it has a  quite large effect at large $s$
(for example, when $s>1.4$GeV). Thus it is demanded that the
polynomial on the $r.h.s.$ of Eq.~(\ref{fS0}) be large enough to
counteract the large negative contribution from the integral when
$s$ is large. It can be shown that the second sheet pole
$f_2(1270)$, at around $M^2\sim 1.63$ GeV$^2$ and $G\sim 0.13$,
provides a negative contribution to the coefficient of $s^2$ term.
That means contributions from other poles are necessary. A virtual
state or a resonance located in some special region can offer a
positive term to the coefficient. For the latter case it can be
verified numerically that, if denoting the 2nd sheet resonance
pole position as $z_0$ on the complex $s$ plane, then
$\mathrm{Re}[z_0]<4m_\pi^2$. On the other hand, a bound state will
offer a negative term, and a third sheet pole contributes through
the right hand cut integral and it also provides a negative
coefficient. Therefore at least one virtual state or one peculiar
second sheet resonance should exist in the I=0 $d$-wave amplitude.
However numerical analysis reveals that the latter case is not
stable,
 the additional resonance will
collapse and convert  to two virtual states. Furthermore, one of
the virtual state moves towards $s=0$ (and hence has a vanishing
effect) and another remains  very close to the location of the
virtual pole when only the latter is used in the fit. Therefore
the conclusion is that a virtual state pole is needed in order to
correctly describe the data. This observation is remarkable, since
the virtual pole's contribution to the phase shift is very small
and hence is hard to be directly seen from the data. On the other
hand the $d$-wave two-loop $\chi$PT amplitude does predict a
virtual state pole at about 1.95 MeV, and the one loop amplitude
at 1.52 MeV using the low energy constants provided by
Ref.~\cite{bijnens}.

Besides the $f_2(1270)$ resonance as described earlier,
Ref.~\cite{PDG04} also provides several other $2^{++}$ resonances
below $1.7GeV$: $f_2(1430)$, $f'_2(1525)$, $f_2(1565)$, except for
the $f_2(1270)$ state. In the present approach they are all
contained in the right hand integral and contribute to $\eta$ as
small dips, and to the phase shift as small kinks. Indeed the
CERN--Munich data\cite{hyams} as shown in fig.~\ref{fitrhc} may
reveal some structure in the region $\sqrt{s}>1.4$GeV,  however
the data are not accurate enough to be useful in distinguishing
those higher mass poles. The data on $\eta$ only exhibits
$f_2(1270)$ clearly, so that fit to $\delta$ and $\eta$ fails to
reproduce other poles unambiguously. Nevertheless these higher
mass poles do play a role in reducing the fit total width of
$f_2^{\mathrm{III}}$, meanwhile they play very little role in
decreasing the total $\chi^2$ and altering any other output. In
the following we perform a 12 parameter fit to the CERN--Munich
data on I=0 $d$ wave $\pi\pi$ scattering phase shift up to 1.6GeV.
The 12 parameters consist of 6 relevant parameters and 6
`irrelevant' parameters. The former are: 1 for the virtual pole
position, 2 for $f_2^{\mathrm {II}}(1270)$ and 3 for $f_2^{\mathrm
{III}}(1270)$ (mass, total width and partial width into $\pi\pi$).
The irrelevant parameters are the total widths and $BR(\pi\pi)$ of
the 3 higher mass states.
 The  extra resonances'
 masses are fixed at their values as provided by
 Ref.~\cite{PDG04}. The fit results are the following,
 \bqa\label{resultvrrhc}
&&\chi^2_{d.o.f.}=96.5/(80-12) \ ;\nonumber\\
&&M_v=1.73\pm 0.04MeV\ ;\nonumber\\
&&M^{\mathrm{II}}_{f_2} =1270.6\pm 3.2MeV\ ,\,\,\,  \Gamma^{\mathrm{II}}_{f_2} =156.5\pm 5.6MeV\ ;\nonumber\\
&&M^{\mathrm{III}}_{f_2} =1278.6\pm 22.2MeV\ ,\,\,\,  \Gamma^{\mathrm{III}}_{f_2}(tot) =225.0\pm 76.2MeV\ ;\nonumber\\
&&BR(\pi\pi)=92.1\pm 2.4\%\ ;\nonumber\\
&&a^0_2=(1.89\pm 0.09)\times 10^{-3}{m^{-4}_\pi}\ ,\,\,\,
b^0_2=(-2.00\pm 0.02 )\times 10^{-4}{m^{-6}_\pi} ;\nonumber\\
 \eqa
 where $a^0_2$ and $b^0_2$ are threshold parameters defined in
 Eq.~(\ref{ab}).
 We also  provide in table~\ref{tab1} a few results of phase shift in
 order to compare with the results from
 Ref.~\cite{BFP74}.
 \begin{table}[bt]%
\centering\vspace{-0.cm}%
\begin{tabular}{|c|c|c|c|c|c|c|c|}%
\hline%
  $M_{\pi\pi}$ (MeV)&400 &600&800&1000&1200&1400&1600  \\%
\hline%
  $\delta^0_2$& 0.125
 & 1.013& 2.914    &10.34
 &44.66  &140.83  &157.85 \\%
\hline%
 $ \triangle\delta^0_2$ & 0.024 &  0.088 &   0.146  &   0.30
&   0.88 &  2.22   &   8.74    \\%
\hline%
 $\eta$&  & $0.968$&$0.955$ &$0.923$&$0.790$&$0.760$ &$0.859$\\%
\hline%
 $\triangle\eta$&  & $0.011$&$0.013$ &$
 0.019$&$0.028$&$0.044$ &$0.064$\\%
\hline%
\end{tabular}%
\caption{\label{tab1}Phase shift value for the I=0,J=2 $\pi\pi$
scattering.
Error bars are obtained using an  error matrix generated by 12 fit parameters.}%
\end{table}%
The results on additional resonances are not included in
Eq.~(\ref{resultvrrhc}), since they totally disagree with their
PDG values.\footnote{ The main reason for the failure is because a
3rd sheet
 narrow resonance's contribution to the phase shift and inelasticity as
 shown by fig.~\ref{kinkanddip} are rather tiny and concentrated in a  small region of
 $\sqrt{s}$ which makes the fit heavily disturbed by the fluctuation of
 data.
 In order to get correct information about these 3rd sheet poles,
 it requires very  dense data with very small error bars.}
 Nevertheless it should be stressed that outputs as
listed above are affected rather little with or without these
additional resonances included. Especially $a^0_2$ is always in
nice agreement with its $\chi$PT value, and $b^0_2$ is not. The
reason why the $b_2^0$ parameter does not agree well with the
$\chi$PT prediction may be that the $b_2^0$ parameter is related
to higher order of momentum expansions and is therefore more
sensitive to distant singularities. A numerical test is made and
it is found that a distant second sheet pole with a very large
width can remedy the discrepancy on $b_2^0$ with very small
influence to other outputs in Eq.~(\ref{resultvrrhc}), but the
situation does not correspond to a local $\chi^2$ minimum and no
meaningful conclusion can be drawn whether there is any physics
behind this discrepancy.

The total $\chi^2$ given by Eq.~(\ref{resultvrrhc}) is not very
good. From fig.~\ref{fitrhc} one sees that the data fluctuation
between 0.8GeV and 1.1GeV contributes a lot to the total $\chi^2$.
It is hard to solely ascribe the discrepancy to the data itself.
Because in the energy region $K\bar K$ threshold opens and should
contribute at least part of the discrepancy. The
Eq.~(\ref{resultvrrhc}) is only a narrow width approximation, and
we know that one condition  for the narrow width approximation to
work is that the energy region is far away from any threshold.
Therefore, the use of Eq.~(\ref{resultvrrhc}) may also contribute
to some of the discrepancy.

To conclude, within an appropriate parametrization form for the
partial wave scattering $S$ matrix incorporated with chiral
symmetry. we have clearly demonstrated that the $f_2(1270)$
resonance exhibits  a twin pole structure, i.e., it generates
simultaneously both a 3rd sheet pole and a 2nd sheet pole. The two
poles locate at similar position, though on different sheets. This
is a  picture exhibited by a typical couple--channel Breit--Wigner
resonance: when the resonance's coupling to the elastic channel is
dominant, two poles on different sheets coexist. In production
experiments one sees the `3rd' sheet pole, whereas in phase shift
analysis  the nearly 180$^\circ$ jump is generated by the second
sheet pole. The pole parameters of $f_2(1270)$ resonance are
determined, the mass and width of the 3rd sheet pole, though
contain rather large error bars, are found in  agreement with the
PDG value.

It is also worth noticing that our approach, for the first time,
uncovers the existence of the virtual state from I=0 $d$-wave
$\pi\pi$ scattering data.  Actually the existence of such a
virtual pole may be proved rigorously without rescuing to
perturbation theory.\footnote{ In fact  there exists a virtual
pole in every partial wave amplitude when $l\geq 2$. Such a
phenomenon leads to the conclusion that there exists a weak
essential singularity at $s=0$ in full amplitude for $\pi^0\pi^0$
scattering, on the second sheet.~\cite{martin}} Our numerical
analysis indicates that the location of the virtual pole is in
good agreement with the $\chi$PT prediction. This is very
remarkable, since the virtual state contributes tiny to the phase
shift and it can hardly be seen directly from the data. The
virtual pole definitely emerges, in the numerical analysis, as
long as left hand cut integral provides a negative contribution
when $s$ is large, hence the conclusion is robust. Our work, to
the best of our knowledge, is the first one in support of the
theoretical prediction by analyzing experimental data.

Combining with our previous analysis on the virtual state in I=2
$s$-wave $\pi\pi$ scattering~\cite{zhou05}, we conclude that the
perturbative chiral expansion in the small $|s|$ region is
trustworthy, at least it provides a self-consistent and
satisfactory description to the data, after embedding into a
correctly unitarized parametrization form. This is not a trivial
statement. For example the Pad\'e approximation  produces a
singularity structure in total disagreement with $\chi$PT
predictions, just not far below the two $\pi$
threshold.~\cite{qin} Therefore it is meaningful  to reexamine
more carefully the structure of the scattering amplitude in the
small $|s|$ region.

{\bf Acknowledgements:} It is a pleasure to thank Zhi-guang ~Xiao
for valuable discussions. This work is supported in part by China
National Nature Science Foundation under grant number 10491306.



\begin{thebibliography}{99}
\bibitem{XZ00}Z.~G.~Xiao and H.~Q.~Zheng, Nucl. Phys. {\bf
A695}(2001)273.
\bibitem{zhou05}Z.~Y.~Zhou $et$ $al.$, JHEP0502(2005)043.
\bibitem{CGL01}G.~Colangelo, J.~Gasser and H.~Leutwyler, Nucl.
Phys. {\bf B603}(2001)125.
\bibitem{LASS}D.~Aston $et$ $al.$ (LASS Collaboration), Nucl. Phys. {\bf B296}(1988)493.
\bibitem{piK}H.~Q.~Zheng $et$ $al.$, Nucl. Phys. {\bf A}733(2004)235.
\bibitem{ang}Q.~Ang $et$ $al.$, Commun. Theor. Phys. {\bf
36}(2001)563.
\bibitem{Ztalk03}H.~Q.~Zheng,
Talk given at International Symposium on Hadron Spectroscopy,
Chiral Symmetry and Relativistic Description of Bound Systems,
Tokyo, Japan, 24-26 Feb 2003, hep-ph/0304173.
\bibitem{XZ05}Z.~G.~Xiao and H.~Q.~Zheng, hep-ph/0502199.
\bibitem{collins}P.~D.~B.~Collins, {\it An Introduction to Regge Theory and High Energy Physics}
 Cambridge University Press,
1977.
\bibitem{PDG04}S.~Eidelman $et$ $al$ (Particle Data Group), Phys.
Lett. B {\bf 592}, 1(2004) and 2005 partial update for edtion
2006(URL: http://pdg.lbl.gov)
\bibitem{bijnens}J.~Bijnens, et al., Phys.Lett. B{\bf
374}(1996);\\
J.~Bijnens,  et al., Nucl. Phys. B{\bf508}(1997)263.
\bibitem{hyams}B.~Hyams $et$ $al.$,
Nucl. Phys. {\bf B} 64(1973)134.
\bibitem{BFP74}J.~L.~Basdevant, C.~D.~Froggatt and J.~L.~Petersen,
Nucl. Phys. {\bf B72}(1974)413.
\bibitem{martin}A.~Martin and F.~Cheung, {\it Analyticity Properties  and Bounds
 of the Scattering Amplitudes}, Gordon and Breach, New York 1970.
\bibitem{qin}G.~Y.~Qin $et$ $al$., Phys. Lett. {\bf B542}(2002)89.



\end{thebibliography}
\end{document}